\documentclass[aps,pra,showpacs,twocolumn,amsmath,amssymb]{revtex4}

\usepackage[english]{babel}
\usepackage{latexsym}
\usepackage{graphics}
\usepackage{epsfig}
\usepackage{color}
\usepackage{url}
\usepackage[normalem]{ulem}

\def\be{\begin{equation}}
\def\ee{\end{equation}}
\def\bea{\begin{eqnarray}}
\def\eea{\end{eqnarray}}
\def\bi{\begin{itemize}}
\def\ei{\end{itemize}}
\def\bin{\begin{enumerate}}
\def\ein{\end{enumerate}}
\def\la{\langle}
\def\ra{\rangle}
\def\ka{\vect{k}}
\def\jeff{J^{\tiny{\mbox{eff}}}_{ij}}

\newcommand{\vect}[1]{\mathbf{#1}}

\newcommand{\cosin}[1]{\cos\left(#1\right)}

\usepackage{graphicx}
\begin{document}
\title{Simulation of frustrated classical XY models with ultra-cold atoms in 3D triangular optical lattices}

%%%%%%%%%%%%%%%%%%%%%%%%%%%%%%%%%%%%%%%%%%%%%%%%%%%%%%%%%%%%%%%%%%%%%%%%%%%%%%%
\author{Arkadiusz Kosior}
\affiliation{
Instytut Fizyki imienia Mariana Smoluchowskiego and
Mark Kac Complex Systems Research Center, 
Uniwersytet Jagiello\'nski, ulica Reymonta 4, PL-30-059 Krak\'ow, Poland}

\author{Krzysztof Sacha}
\affiliation{
Instytut Fizyki imienia Mariana Smoluchowskiego and
Mark Kac Complex Systems Research Center, 
Uniwersytet Jagiello\'nski, ulica Reymonta 4, PL-30-059 Krak\'ow, Poland}

\date{\today}

\begin{abstract}
Miscellaneous magnetic systems are being recently intensively investigated because of their potential  applications in modern  technologies. Nonetheless, a many body dynamical description of complex magnetic systems may be cumbersome, especially when the system exhibits a geometrical frustration. This paper deals with simulations of the classical XY model on a three dimensional triangular lattice with anisotropic couplings, including an analysis of the phase diagram and a Bogoliubov description of the dynamical stability of  mean-field stationary solutions. We also discuss the possibilities of the realization of Bose-Hubbard models with complex tunneling amplitudes in shaken optical lattices without breaking the generalized time-reversal symmetry and the opposite, i.e. real tunneling amplitudes in systems with the time-reversal symmetry broken.
\end{abstract}


\maketitle
%%%%%%%%%%%%%%%%%%%%%%%%%%%%%%%%%%%%%%%%%%%%%%%%%%%%%%%%%%%%%%%%%%%%%%%%%%%%%%%

%-------------------------------------------------------------------------------------------------------------------------------
\section{Introduction}
\label{seci}
%-------------------------------------------------------------------------------------------------------------------------------
This paper was inspired by an experiment of the Hamburg University group who managed to build the first experimental realization of the two-dimensional (2D) quantum simulator of classical frustrated magnetism in an ultra-cold atomic gas in a triangular optical lattice \cite{struck2011,Olschlager, eckardt2010}. Frustrated magnetism is the result of competition between interactions and geometry of a lattice and constitutes a very active field of research \cite{morris09,fennell09,balents10}. 

Quantum simulators are easily controllable quantum systems that can be employed to mimic others \cite{QuantumSimulators}. If considered systems are similar enough, one may be able to map one system's Hamiltonian onto the Hamiltonian of the simulator. Consequently, a time evolution of the quantum simulator imitates a simulated one. Quantum simulators are extremely useful when issues in consideration are substantially too troublesome for computers or their direct observation is laborious or hardly possible.

An exquisite example of a quantum simulator is a system of ultra-cold bosonic gas in an optical lattice \cite{OL_2005, lewen07, OL_2008, OL_2012}. Among others, that is because of a great flexibility of optical lattices, as their geometry can be changed easily and a possibility to manipulate atomic interactions by changing depth of a lattice or via Feshbach resonances \cite{feschbach1, feschbach2}. An optical lattice has a structure of an ideal crystal, being devoid of defects present in material crystals, and therefore is a natural simulator of many solid state physics problems, such as miscellaneous spin models.

In our work, we simulate the classical XY model with nearest neighbor interactions characterized by $J_{ij}$ couplings 
\be
H= - \sum_{\la ij\ra}J_{ij} \;\vec{S}_{\vect{r}_i}\cdot \vec{S}_{\vect{r}_j},
\ee
where
\be
\vec{S}_{\vect{r}_i} = \left( \cos\theta_{\vect{r}_i},\sin \theta_{\vect{r}_i}  \right),
\label{spinangle}
\ee 
is a two component classical vector representing a direction of a spin on $i$-th lattice site. Simulations of such a model in ultra-cold atomic gases are possible due to an identification of the directional angle $\theta_{\vect{r}_i}$ with a local phase of bosonic condensate wavefunction $\theta_{\vect{r}_i}=\vect{k}\cdot \vect{r}_i$ of atoms in an optical lattice of the same geometry as the spin lattice.  The vector $\vect{k}$ is a quasi-momentum vector of a single atom and the couplings $J_{ij}$ are related to elements of tunneling amplitude matrix. By doing so, an energy per a single spin can be associated with a dispersion relation of an atom of an ideal non-interacting gas. A key to simulations is a possibility of manipulating $J_{ij}$ values. One may conclude that it is impossible to simulate anti-ferromagnetism since tunneling amplitudes $J_{ij}$ naturally have to be non-negative. However, it turns out that they can be effectively made negative or even complex \cite{eckardt2005,eckardt2010,struck2011,TRSbreaking,spielman,struck2012,struck2012a}.

\begin{figure}
\begin{center}
\resizebox{0.95\columnwidth}{!}{\includegraphics{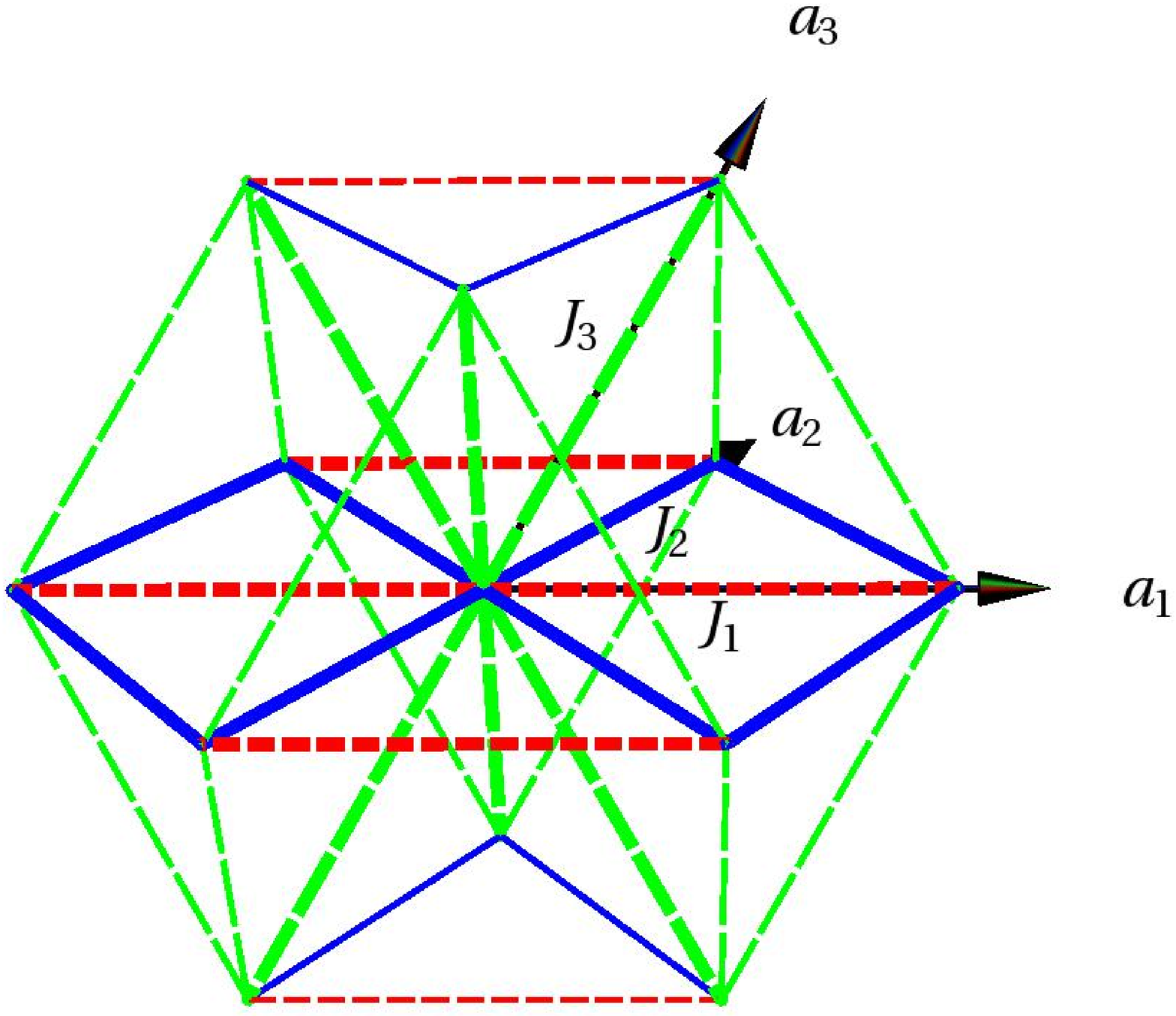}} 
\caption{(Color on line) Triangular optical lattice corresponding to the trigonal  Bravais lattice with anisotropic tunneling amplitudes. Lattice vectors $\vect{a}_i$ form a regular tetrahedron. Amplitudes $J_i$ related to tunneling along different directions are indicated in the figure, i.e. $J_1$ corresponds to the tunneling along dashed (red) lines, $J_2$ along solid (blue) lines, and $J_3$ along long-dashed (green) ones.}
\label{int}
\end{center}
\end{figure}

A model in our interest is the classical XY model on a 3D triangular lattice corresponding to the trigonal structure \cite{solidstate} with basis vectors forming equal angles of $\pi/3$ and anisotropic tunneling amplitudes. The primitive vectors of the considered Bravais lattice read
\bea
\vect{a}_1&=&a\;\vect{e}_x, \cr&& \cr
\vect{a}_2&=&\frac{a}{2}\;\left(\vect{e}_x+\sqrt{3}\;\vect{e}_y\right), \cr&& \cr
\vect{a}_3&=&\frac{a}{2\sqrt{3}}\;\left(\sqrt{3}\;\vect{e}_x+\vect{e}_y+2\sqrt{2}\;\vect{e}_z\right),
\label{a1a2a3}
\eea
where $a$ is the lattice constant. The vectors $\vect{a}_1,\vect{a}_2,\vect{a}_3$ form a regular tetrahedron, see Fig. \ref{int}.

 In the present publication we mainly focus on real tunneling amplitudes which have three different values depending on a tunneling direction as depicted in Fig.~\ref{int}. Following the preceding identification, the energy per spin depends only on one vector $\vect{k}$ and adopts a form 
\be
\label{disp}
\begin{split}
E(\ka) & = -2\biggl\{J_1 \cosin{\ka \cdot \vect{a}_1}+J_2 \Bigl[\cosin{\ka \cdot \vect{a}_2}+\\
 & +\cosin{\ka \cdot(\vect{a}_2-\vect{a}_1})\Bigr]+J_3 \Bigl[\cosin{\ka \cdot \vect{a}_3}+\\
& + \cosin{\ka \cdot(\vect{a}_3-\vect{a}_1})+\cosin{\ka \cdot(\vect{a}_3-\vect{a}_2)}\Bigr]\biggr\}.
\end{split}
\ee

This paper is organized as follows. In Sec.~\ref{secii} we present a background on how to make simulations of classical frustrated magnetism possible. Firstly, we propose a way of constructing a 3D triangular lattice. Then, to manipulate $J_{ij}$ we revise a procedure proposed in Refs.~\cite{eckardt2005,eckardt2010} to apply it to our model. In Sec.~\ref{seciii} we describe the results of our simulations, i.e. a phase diagram and phase transitions. In Sec.~\ref{seciv} we analyze the stability of stationary states within the Bogoliubov formalism.   Section~\ref{timereversal} is devoted to a discussion of the relation between complex tunneling amplitudes in effective Bose-Hubbard models and the presence or absence of the  generalized time-reversal symmetry of systems. Especially we show that complex amplitudes can be realized by the shaking of an optical lattice that does not break the generalized  time-reversal symmetry. Finally, in Sec.~\ref{concl} we conclude.

%-------------------------------------------------------------------------------------------------------------------------------
\section{Construction of 3D triangular lattice model}
\label{secii}
%-------------------------------------------------------------------------------------------------------------------------------
%-------------------------------------------------------------------------------------------------------------------------------

%-------------------------------------------------------------------------------------------------------------------------------
%-------------------------------------------------------------------------------------------------------------------------------
\subsection{Experimental setup}
\label{seciia}
\begin{figure}
\begin{center}
\resizebox{0.45\columnwidth}{!}{\includegraphics{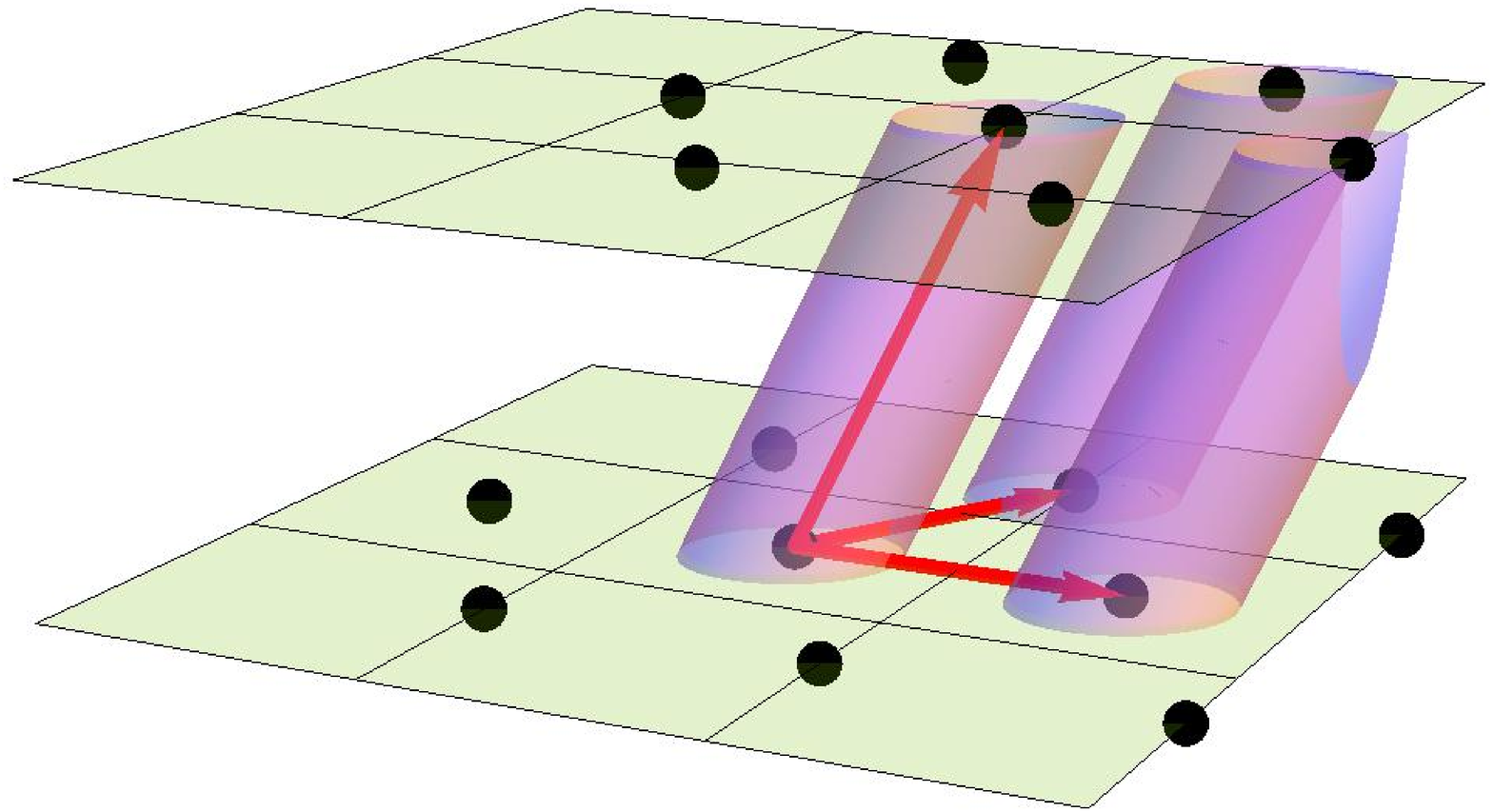}} 
\resizebox{0.45\columnwidth}{!}{\includegraphics{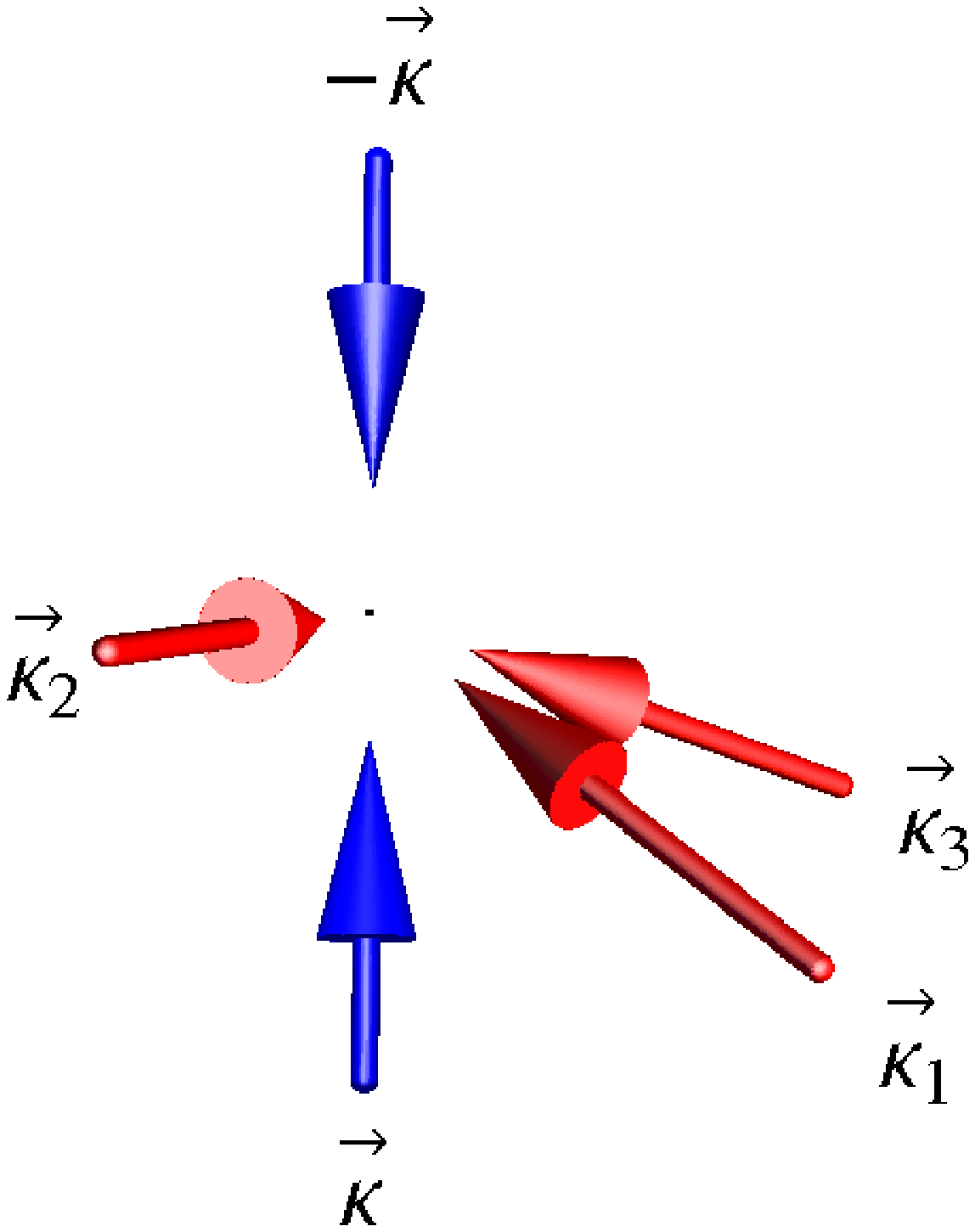}} 
\caption{(Color on line) Left panel illustrates the idea of realization of a 3D triangular optical lattice by cutting a 2D triangular lattice consisting of 1D tubes. Arrows indicate primitive vectors of the resulting 3D lattice. Right panel shows configuration of laser beams that create a 3D triangular optical lattice. Wave vectors $\vec{\kappa}_i$ and $\vec{\kappa}$ of the laser beams correspond to Eq.~(\ref{beams}).}
\label{lasery}
\end{center}
\end{figure}

%-------------------------------------------------------------------------------------------------------------------------------
%-------------------------------------------------------------------------------------------------------------------------------
Optical lattices used to be formed by a system of uni-polarized counter-propagating laser beams that form a standing-wave. As practical as it is for quasi-cubical lattices, it is hardly useful when dealing with triangular ones. The interfering running waves can be shifted to coincide at any angle. In the case of 2D triangular lattice three coplanar red-detuned and uni-polarized beams should be pointed inwards forming an angle of~$2\pi/3$ with each other. Such a construction was successfully realized experimentally in Refs.~\cite{becker10,struck2011}. This method could be further generalized into three-dimensions using four beams, however there emerges a problem of finding their proper polarizations. Under this circumstances, in order to construct a 3D triangular lattice, we propose to combine both methods. Namely, we first construct a 2D lattice, which is in fact a lattice of one-dimensional tubes \cite{becker10,struck2011}. Then, we add an extra oblique pair of counter-propagating beams that form a standing-wave. We choose, in a Cartesian basis, wave vectors of running waves (vectors $\vec{\kappa}_1,\vec{\kappa}_2,\vec{\kappa}_3$) and beams forming a standing wave (vectors $\pm\vec{\kappa}$) to be
\begin{gather}
\vec{\kappa}_1 = \frac{\kappa_L}{\sqrt{6}}
\begin{pmatrix}
- \sqrt{3} \\
-1\\
\sqrt{2}
\end{pmatrix},
\quad
\vec{\kappa}_2 = \frac{\kappa_L}{3 \sqrt{6}}
\begin{pmatrix}
\sqrt{3} \\
-7 \\
\sqrt{2}
\end{pmatrix},\nonumber
\\
\vec{\kappa}_3 = \frac{\kappa_L}{3\sqrt{6}}
\begin{pmatrix}
-3\sqrt{3} \\
5 \\
\sqrt{2}
\end{pmatrix},
\quad
\vec{\kappa} = \kappa_L'
\begin{pmatrix}
0 \\
0 \\
1
\end{pmatrix},
\label{beams}
\end{gather}
where $\kappa_L$, $\kappa_L'$ are moduli of the wave vectors and $\kappa_L'=\kappa_L/\sqrt{3}$.  The coordinate frame was chosen accordingly to construct the lattice given by the primitive vectors (\ref{a1a2a3}), where we imposed a condition $a=3/(\sqrt{2}\kappa_L)$ relating $\kappa_L$ and the lattice constant $a$. These proceedings are illustrated in Fig.~\ref{lasery}.

%-------------------------------------------------------------------------------------------------------------------------------
%-------------------------------------------------------------------------------------------------------------------------------
\subsection{Effective description of driven lattices}
\label{seciib}
%-------------------------------------------------------------------------------------------------------------------------------
%-------------------------------------------------------------------------------------------------------------------------------
As it has been already sketched, the simulations are based on  manipulations of tunneling amplitudes.
It can be done without changing the lattice constant. An ingenious idea, proposed by Eckardt, Weiss, and Holthaus \cite{eckardt2005}, is to manipulate the values of tunneling matrix elements by introducing a periodic lattice modulation. Experimentally the motion of the lattice can be induced by a periodic frequency difference between any two interfering beams. In result, one can make the lattice follow any desired trajectory $\vect{R}(t)$.

An ultra-cold bosonic gas in a driven lattice is described in the co-moving frame by an explicitly time dependent Hamiltonian of the Bose-Hubbard form 
\bea\label{bosehubbard}
\hat{H}_{BH}(t)  &=& -\sum_{\langle i,j \rangle}J_{ij} \hat{b}_i^\dagger\hat{b}_j  + \frac{U}{2}\sum_i \hat{n}_i (\hat{n}_i -1) \cr
& & -\sum_i \hat{n}_i \;\vect{r}_i \cdot \vect{F}(t),
\eea
where $U$ is on site interaction energy and $\vect{F}(t) = -m\ddot{\vect{R}}(t) $ is an inertial force \cite{eckardt2005}. We assume here that the shaking of the lattice does not lead us out of the lowest energy band of a lattice problem. It is valid provided that energy scale related to the driving frequency $\omega$ is considerably lower than the energy gap between the bands \cite{Holthaus_2012}. 

Although energy eigenvalues do not exist for time dependent systems, an ultra-cold bosonic gas in a periodically driven lattice does not constitute a non-equilibrium system. It was proven that every solution of the periodically time dependent Schr\"{o}dinger equation can be unambiguously expressed by the so-called Floquet states \cite{Holthaus_2012, floquet1,floquet2,floquet3}. That is the reason why the periodicity is of vital importance here. 

The term $\sum_i \hat{n}_i \vect{r}_i \cdot \vect{F}(t)$ can be eliminated from the Hamiltonian (\ref{bosehubbard}) with the help of the unitary transformation 
\be
U(t)=\exp \left( -\frac{i  }{\hbar} \sum_i \hat{n}_i W_{i}(t)\right),
\label{utrans}
\ee
with
\be\label{wi}
W_i(t)=-\vect{r}_i\cdot \int_0^tdt'\; \vect{F}(t'). 
%-\frac{1}{T}\int_0^Tdt''\int_0^{t''}dt'\vect{F}(t')\right),
\ee
The resulting Hamiltonian has time dependent tunneling amplitudes
\be
J'_{ij}(t)= J_{ij} \exp \left( \frac{i  }{\hbar} \,W_{ij}(t) \right),\quad W_{ij}(t)=W_i(t)-W_j(t).
\ee
If we assume now a high frequency regime, i.e. when $\hbar \omega$ is considerably larger than all energy scales in the Hamiltonian, then we can time average fast oscillating phases and our system can be described by the effective time independent Bose-Hubbard Hamiltonian \cite{eckardt2005,Holthaus_2012}
\bea
\hat{H}  &=& -\sum_{\langle i,j \rangle}J_{ij}^{\rm eff}\; \hat{b}_i^\dagger\hat{b}_j  + \frac{U}{2}\sum_i \hat{n}_i (\hat{n}_i -1).
\label{BHH}
\eea
with effective, renormalized tunneling amplitudes 
\be
\label{jeff}
\jeff=\frac{1}{T}\int_0^Tdt\;J'_{ij}(t),
\ee
where $T=2\pi/\omega$. Such effective models have been realized in several experiments \cite{lignier07,kierig08,zenesini09,struck2011,struck2012}.

%-------------------------------------------------------------------------------------------------------------------------------
%-------------------------------------------------------------------------------------------------------------------------------
\subsection{Effective tunneling amplitudes for 3D triangular lattice}
\label{seciic}
%-------------------------------------------------------------------------------------------------------------------------------
%-------------------------------------------------------------------------------------------------------------------------------

\begin{figure}
\begin{center}
\resizebox{0.9\columnwidth}{!}{\includegraphics{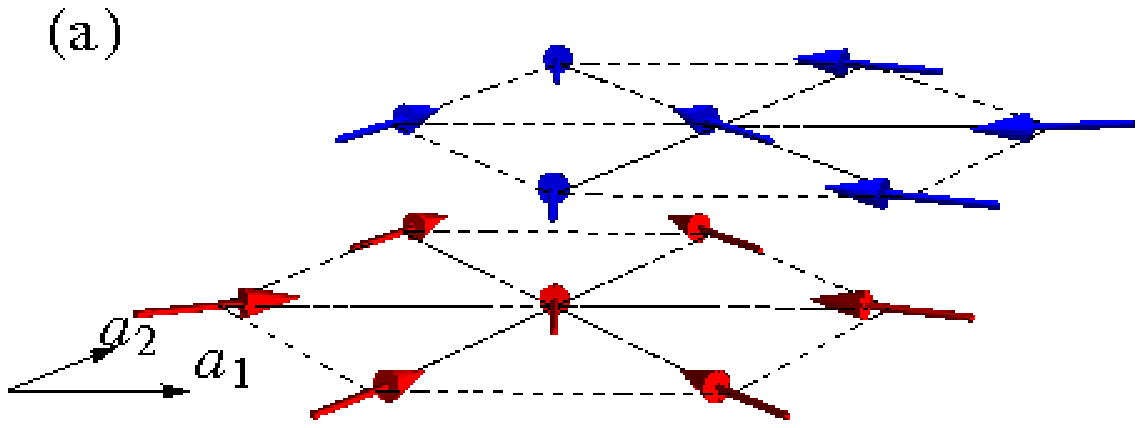}} 
\resizebox{0.9\columnwidth}{!}{\includegraphics{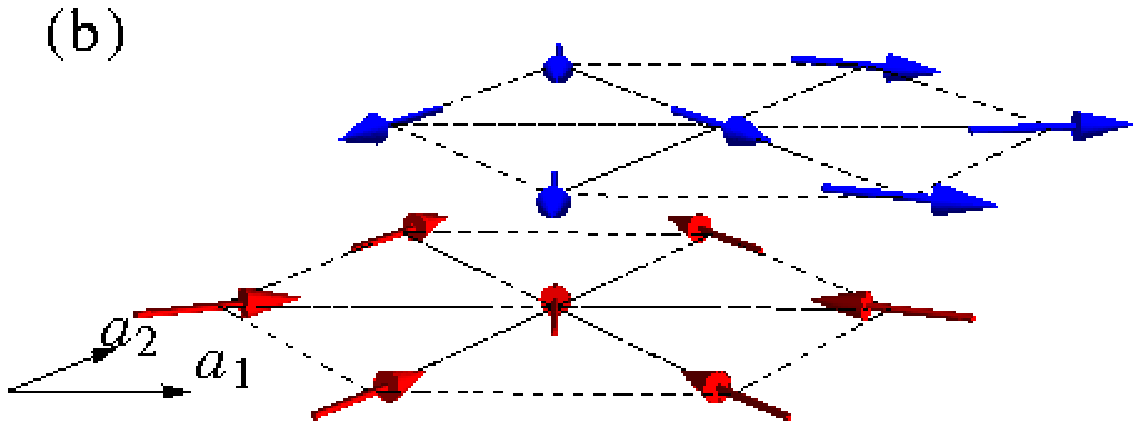}} 
\caption{(Color on line) Panel (a) shows a configuration of spins in a 3D triangular lattice corresponding to the ground state of the system with tunneling amplitudes: $J_1=-0.3J$, $J_2= 0.2J$ and $J_3= 0.3J$. It belongs to a spiral phase indicated in Fig.~\ref{diagram} as S2. By rotating spin vectors (\ref{spinangle}) in every second $\vect{a}_1\vect{a}_2$ plane by $\pi$ we obtain the configuration shown in panel~(b) which corresponds to the ground state of the system where $J_3\rightarrow-J_3$. The change of the sign of the $J_3$ coupling changes the corresponding interactions from ferromagnetic to anti-ferromagnetic what, in turn, requires the rotation of the spins in order to obtain a ground state configuration.}
\label{symmetry}
\end{center}
\end{figure}

We consider a 3D triangular lattice corresponding to the trigonal Bravais lattice with primitive vectors forming a regular tetrahedron that can be created by means of laser beams.

In the absence of particle interactions ($U=0$), energy eigenvalues of the Hamiltonian (\ref{BHH}) correspond to the dispersion relation of a single atom in the presence of a lattice potential. In the presence of the repulsive interactions ($U> 0$) and for a homogeneous system (i.e. $\la \hat n_i\ra=n$ where $n$ is the mean number of atoms per lattice site), the ground state of the Gross-Pitaevskii equation \cite{pethick} is related to the minimum of the dispersion relation. Now we would like to show how to realize cold atom experiments where the dispersion relation of a single particle in the 3D triangular lattice is given by Eq.~(\ref{disp}) and the tunneling amplitudes $J_1$, $J_2$ and $J_3$ can be controlled experimentally.

When $J_3$ coupling vanishes, the model is equivalent to two-dimensional one, which was already investigated \cite{struck2011,Olschlager, eckardt2010}. In the 2D case, the shaking of the lattice on an elliptical trajectory was used to manipulate the tunneling amplitudes $J_1$ and $J_2$ \cite{struck2011}. It turns out that in the 3D case, changing the orientation of the ellipse and its shape allows us also to manipulate  all three amplitudes. We assume that the triangular lattice is moving on an elliptical trajectory with the velocity
\bea
\dot{\vect{R}}(t)&=\frac{1}{m\omega}\left[F_x\vect{e}_x\cos(\omega t)-\left(F_y\vect{e}_y+F_z\vect{e}_z\right)\sin(\omega t)\right]\theta(t), \cr &&
\label{veloc}
\eea
where the Heaviside step function $\theta(t)$ indicates that the shaking of the lattice begins at $t=0$. Consequently, the inertial force $\vect{F}(t) = -m\ddot{\vect{R}}(t)$ contains Dirac-delta contribution which has to be taken into account \cite{Holthaus_2012}. In the case of the 3D triangular lattice under consideration, if we want to have only three different tunneling amplitudes as indicated in Fig.~\ref{int}, the $z$-component of the inertial force has to fulfil
\be
F_z=\frac{F_y^2-F_x^2}{4\sqrt{2}F_y}.
\ee
Then, the three tunneling amplitudes read
\bea
J_1&=&J\;{\cal J}_0\left(\frac{F_xa}{\hbar \omega}\right), \cr &&\cr
J_2&=&J\;{\cal J}_0\left(\frac{a}{2\hbar\omega}\sqrt{F_x^2+3F_y^2}\right), \cr &&\cr
J_3&=&J\;{\cal J}_0\left(\frac{a}{4\sqrt{3}\hbar\omega}\frac{F_x^2+F_y^2}{F_y}\right), 
\label{j1j2j3}
\eea
where $J$ is the tunneling rate of atoms between neighboring sites of the static (not shaken) triangular optical lattice and ${\cal J}_0$ represents the Bessel function of the zero order. The oscillatory character of the Bessel function allows us to change values and signs of the tunneling amplitudes by changing the parameters of the shaking, i.e. $F_x$, $F_y$ and $a/\omega$.

In Eq.~(\ref{veloc}) we have chosen a particular direction of the velocity vector at the moment we turn the shaking on. By changing $\omega t\rightarrow \omega t +\varphi$, one can choose a different initial direction. However, we would like to stress that the phase $\varphi$ is irrelevant and the tunneling amplitudes in Eq.~(\ref{j1j2j3}) are the same independently of the choice of $\varphi$. 

We have applied the unitary transformation (\ref{utrans}) where the inertial force corresponds to the elliptical trajectory defined in Eq.~(\ref{veloc}) and the resulting tunneling amplitudes (\ref{j1j2j3}) are real valued. If we apply slightly different unitary transformation, i.e. with $W_i(t)\rightarrow W_i(t)+\gamma_i$, where $\gamma_i$'s are arbitrary real numbers, the resulting effective Hamiltonian has also a form of the Bose-Hubbard Hamiltonian (\ref{BHH}) but tunneling amplitudes may become complex valued. The freedom in choice of $\gamma_i$ (the so-called gauge freedom \cite{struck2012}) is related to the freedom in choice of the global phase of the Wannier function localized at that site. Eigenstates of the system are obviously the same regardless of the choice of $\gamma_i$'s --- for different sets of $\gamma_i$'s they are written in different basis only. In experiments momentum distributions of atoms corresponding to ground states are measured. Momenta of atoms are detected in the laboratory frame. The Hamiltonian (\ref{bosehubbard}) is related to the frame co-moving with an optical lattice but the unitary transformation (\ref{utrans}), where $\vect{F}(t) = -m\ddot{\vect{R}}(t)$ and $\dot{\vect{R}}(t)$ is given in Eq.~(\ref{veloc}), leads to the frame where the momentum distributions coincide with those measured in the laboratory frame in experiments \cite{Holthaus_2012}. Therefore, the choice of the same $\gamma_i$ for all lattice sites (e.g.  $\gamma_i=0$) that results in the Bose-Hubbard Hamiltonian with the effective tunneling amplitudes (\ref{j1j2j3}) is the unique choice justified by the measurement procedure.

%-------------------------------------------------------------------------------------------------------------------------------
%-------------------------------------------------------------------------------------------------------------------------------
\section{The results of simulation }
\label{seciii}
%-------------------------------------------------------------------------------------------------------------------------------
%------------------------------

%-------------------------------------------------------------------------------------------------------------------------------
%-------------------------------------------------------------------------------------------------------------------------------
\subsection{Phase diagram and phase transitions}
\label{seciiia}
%-------------------------------------------------------------------------------------------------------------------------------
%-------------------------------------------------------------------------------------------------------------------------------

\begin{figure}
\begin{center}
\resizebox{0.9\columnwidth}{!}{\includegraphics{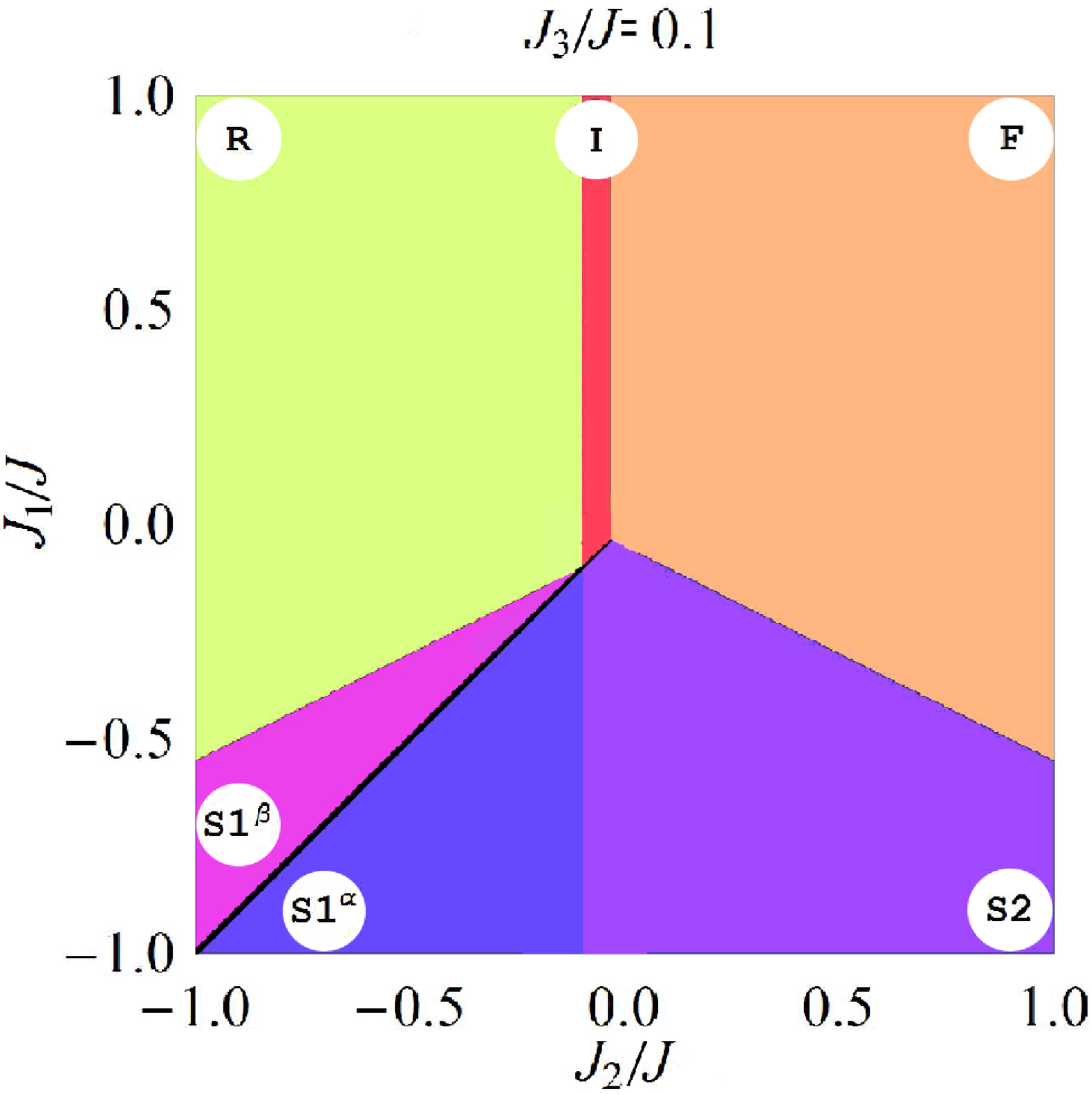}} 
\resizebox{0.9\columnwidth}{!}{\includegraphics{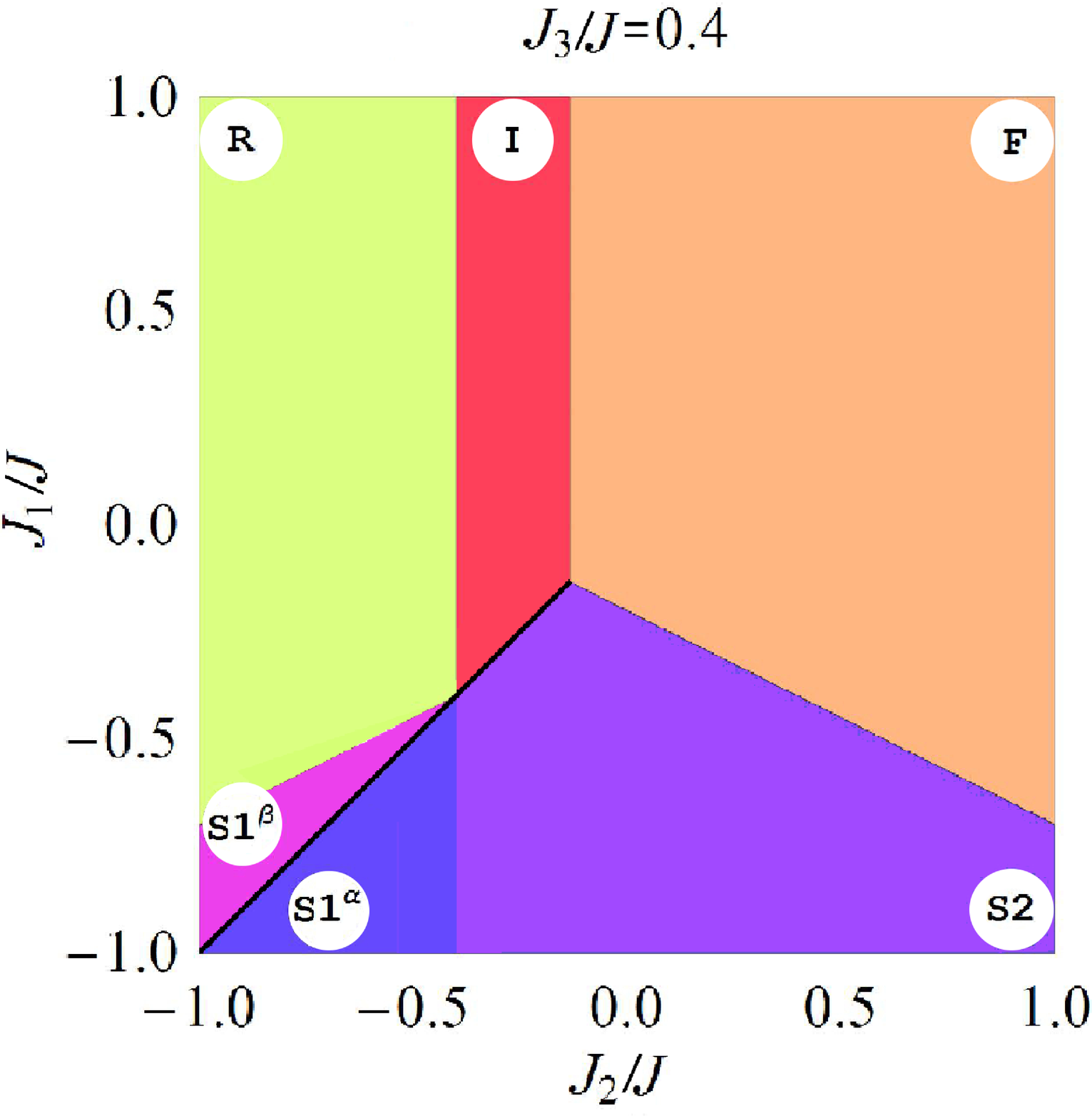}} 
\caption{(Color on line) Phase diagrams for two different positive values of $J_3$, i.e. $J_3=0.1J$ (top panel) and $J_3=0.4J$ (bottom panel). Parameter regions of ferromagnetic and rhomboidal phases are denoted by F and R, respectively. Between these regions there is an intermediate phase denoted by I. There are three regions of different spiral phases: S2, S$1^\alpha$ and S$1^\beta$. Transitions between all phases are continues except the transitions between S$1^\alpha$ and S$1^\beta$ and between S2 and I.  }
\label{diagram}
\end{center}
\end{figure}

\begin{figure}
\begin{center}
\resizebox{0.3\columnwidth}{!}{\includegraphics{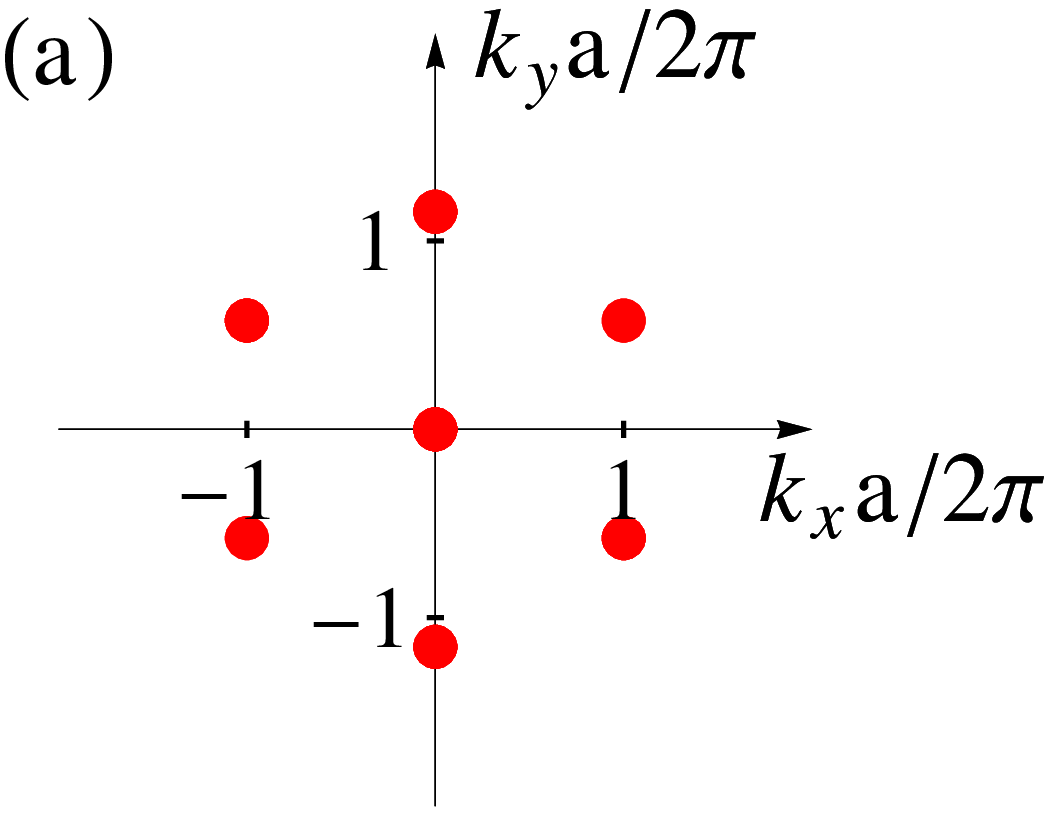}} 
\resizebox{0.3\columnwidth}{!}{\includegraphics{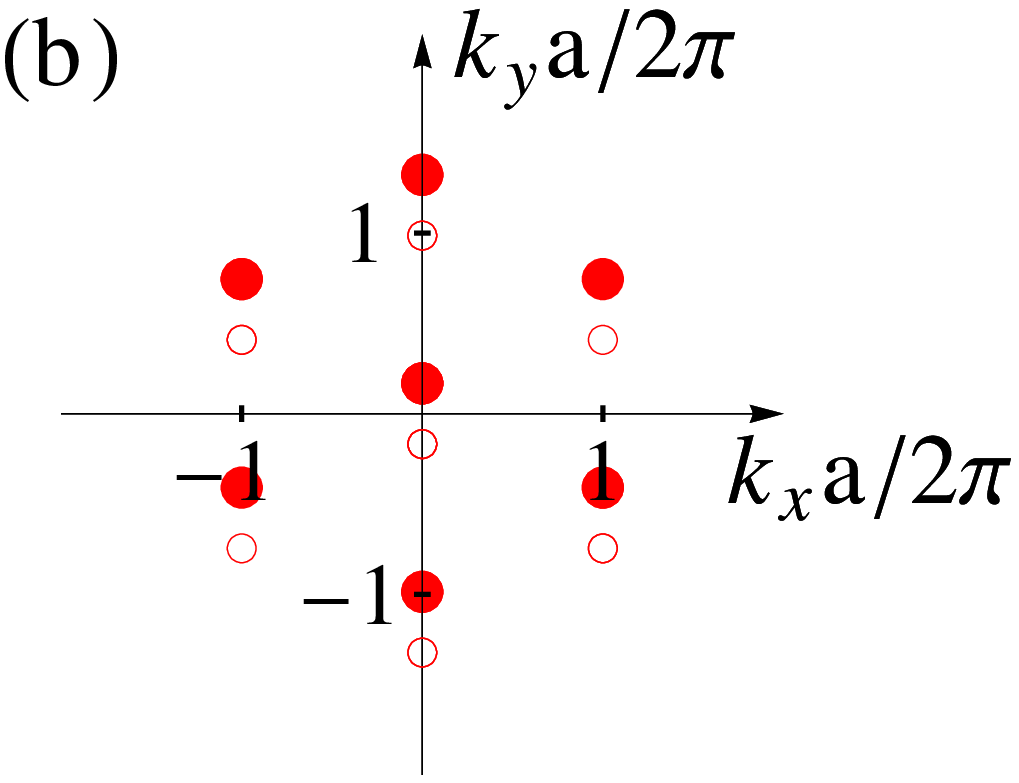}} 
\resizebox{0.3\columnwidth}{!}{\includegraphics{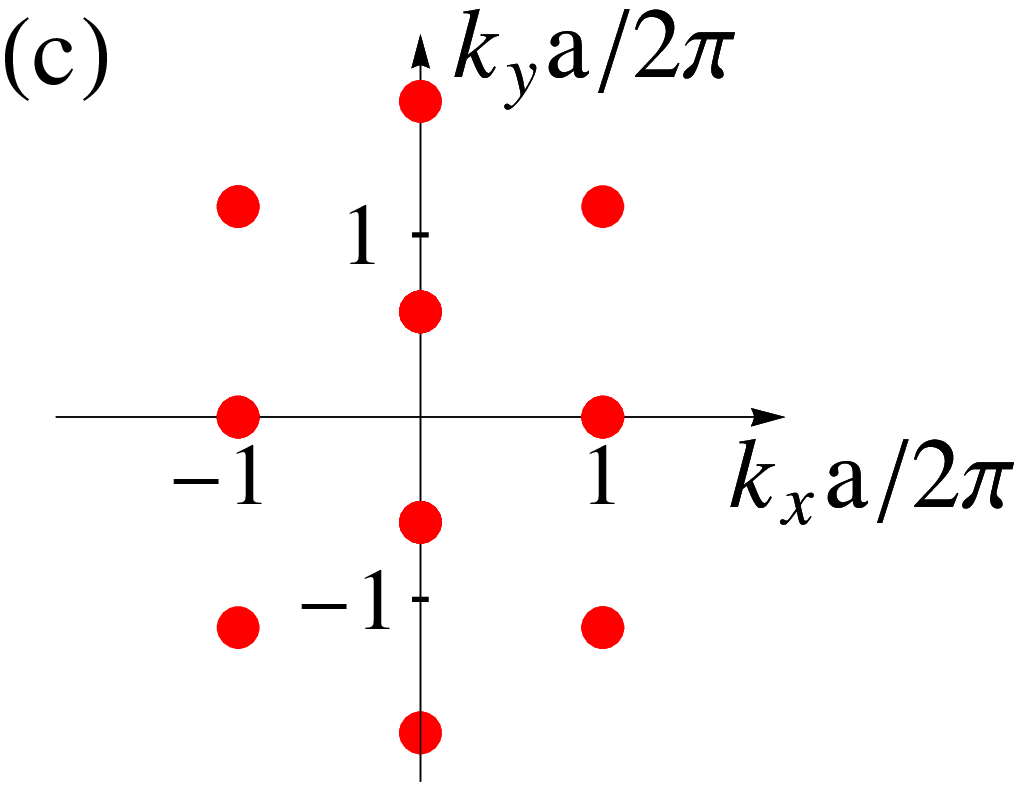}} 
\caption{(Color on line) Projection (along $z$-axis) of momentum distributions of atoms prepared in ground states of a 3D triangular optical lattice. Panels (a) and (c) are related to the ferromagnetic and rhomboidal phases, respectively. Panel~(b) shows two degenerate intermediate states (one is depicted by full symbols and the other by open ones) which allow us to pass continuously from the ferromagnetic phase to the rhomboidal one. Tunneling amplitudes are the following $J_1=0.2J$, $J_3=0.3J$ and $J_2=-0.1J$ (a), $J_2=-0.11J$ (b) and $J_2=-0.3J$ (c).}
\label{intermediate}
\end{center}
\end{figure}

In this subsection we present a description of a phase diagram and phase transitions of the model with real coupling constants, expressed by Eq.~(\ref{j1j2j3}). In our study not only did we obtain ground state configurations which can be considered as generalizations of 2D ones but also new magnetic phases. In addition, we observe some interesting phenomena at phase transition boundaries. We will skip a detailed description of ground state phases of the system that are similar to the 2D case \cite{eckardt2010,struck2011,Olschlager} and rather concentrate on new phenomena that are present in the 3D lattice only. 

A ground state solution of the Gross-Pitaevskii equation of a homogeneous atomic gas in a 3D triangular lattice corresponds to a minimum of the dispersion relation (\ref{disp}).
The appearance of the $J_3$ coupling implies some novelties as compared to the 2D case. First of all, every ground state configuration in the 2D case has two 3D counterparts. The latter can be transformed into one another by rotating spins (\ref{spinangle}) in every second $\vect{a}_1\vect{a}_2$ plane by $\pi$ --- Fig.~\ref{symmetry} presents an example. It is due to a symmetry of the dispersion relation (\ref{disp}) which remains unchanged under a simultaneous replacement $J_3 \rightarrow - J_3$ and $\vect{k}\rightarrow \vect{k} +\vect{b}_3/2$ where $\vect{b}_3$ is a primitive vector in the reciprocal space that is orthogonal to the $\vect{a}_1\vect{a}_2$ planes. Because of this symmetry, in discussing the phase diagram it is sufficient to restrict  ourselves to positive values of the $J_3$ coupling only. Phase diagrams for two different positive values of $J_3$ are presented in Fig.~\ref{diagram}.

In the 2D case by changing $J_2$ we can switch from ferromagnetic (F) to rhomboidal (R) phases and this transition is the first order transition \cite{eckardt2010,struck2011}. In the 3D case the transition is no longer discontinuous because a new intermediate (I) phase emerges. Spins can now reorganize smoothly from the ferromagnetic configuration to the rhomboidal one as illustrated in Fig.~\ref{intermediate}.  It is a pure consequence of the geometrical frustration of the system. Namely, a competition between $J_3$ and $J_2$ couplings enables intermediate configurations.

 Back in two-dimensions, spiral configurations S1 and S2 are the only ones that exhibit signs of frustrations \cite{eckardt2010,struck2011}. In 2D there is a continuous transition between these phases and they are separated by a boundary phase of anti-ferromagnetic chains. Naturally, this boundary appears when $J_2$ vanishes while $J_1$ stays negative, i.e. where anti-ferromagnetic chains can be formed. In the 3D system, because of the extra interaction, the corresponding boundary is shifted. Anti-ferromagnetic chains do emerge in 3D as well, but the chains are not independent because there are correlations between spins belonging to different chains.

\begin{figure}
\begin{center}
\resizebox{0.8\columnwidth}{!}{\includegraphics{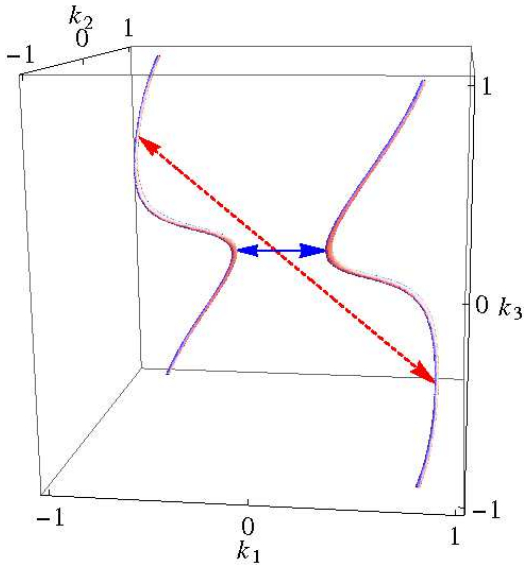}} 
\caption{(Color on line) Curves form a set of quasi-momenta $\vect{k}$ corresponding to the ground state energy of the system at the border between S1$^\alpha$ and S1$^\beta$ phases, i.e. for $J_1=J_2=-0.15J$ and $J_3=0.1J$. The transition between the phases is discontinuous. 
 Starting from the S1$^\alpha$ phase and passing the boundary, the ground state of the system jumps from one of the points indicated by the red (dash) vectors to one of the points indicated by the blue (solid) vectors,  which  represent two degenerate ground states of the S1$^\beta$ phase. Note that in the figure we do not use the Cartesian coordinate frame but the frame corresponding to the primitive vectors of the reciprocal space.} 
\label{discont}
\end{center}
\end{figure}

What is curious in 3D, the region of the S1 spiral configuration of the 2D model is now divided into two regimes that we denote by S1$^\alpha$ and S1$^\beta$. The transition between these new spiral phases is of the first order (even for infinitesimal $J_3$ values) and it goes through a boundary configuration with non-trivial degeneracy which is illustrated in Fig.~\ref{discont}. When we start with the S1$^\alpha$ phase, change the parameters and reach the boundary we end with a different ground state than when we do the same but start with the S1$^\beta$ phase, see Fig.~\ref{discont}.
A Similar phenomenon takes place on the boundary between the S2 phase and the intermediate phase I, i.e. this transition is also discontinuous.

All phases presented in Fig.~\ref{diagram} are doubly degenerated except F and R. The reason for the degeneracy is the invariance of the dispersion relation (\ref{disp}) under reflection in the quasi-momentum space. That is, by changing $\vect{k}\rightarrow-\vect{k}$ in a given ground state one obtains another ground state. In the case of the F and R phases such a transformation does not produces a new ground state because it reduces to a translation by a primitive vector of the reciprocal space and thus leads to the same state, see Fig.~\ref{intermediate} 

%-------------------------------------------------------------------------------------------------------------------------------
%-------------------------------------------------------------------------------------------------------------------------------

\subsection{Bose-Hubbard model with complex tunneling amplitudes}
\label{seciiib}
%-------------------------------------------------------------------------------------------------------------------------------

We have discussed the Bose-Hubbard model (\ref{BHH}) with effective time-independent tunneling amplitudes $J_{ij}^{\rm eff}$ which are real valued. However, it is also possible to realize complex $J_{ij}^{\rm eff}$. Usually it is done by applying the shaking of an optical lattice that breaks the time-reversal symmetry \cite{TRSbreaking,struck2012} (see also \cite{spielman} for a different method) but it turns out that one can realize complex $J_{ij}^{\rm eff}$ even in the presence of this symmetry --- see Sec.~\ref{timereversal}.

The introduction of complex coefficients modifies the dispersion relation (\ref{disp}) and shifts its extrema. This modification reduces to a replacement of all couplings $J_i$ in (\ref{disp}) with their absolute values $|J_i|$ and shifting arguments of the corresponding cosines by their complex phase $\varphi_i$, i.e. $J_i \cosin{\ldots}\longrightarrow |J_i| \cosin{\ldots -\varphi_i}$. 
While $\varphi_3$ can be trivially eliminated with the help of a translation $\ka \rightarrow \ka - \frac{\varphi_3}{2\pi}\vect{b}_3$  ($\vect{b}_3$ is a primitive vector in the reciprocal space), the remaining phases $\varphi_1,\,\varphi_2$ change structure of the dispersion relation essentially. In particular, the quasi-momentum-reversal symmetry, whose presence is the reason for the double degeneracy of the spiral and intermediate phases discussed in Sec.~\ref{seciiia}, is broken. Furthermore,  complex couplings make it possible to take a detour around any phase boundary and consequently avoid any first order phase transition present in the real coupling case.

%-------------
%-------------------------------------------------------------------------------------------------------------------------------
\section{Stability of stationary states: Bogoliubov analysis}
\label{seciv}
%-------------------------------------------------------------------------------------------------------------------------------

\begin{figure}
\begin{center}
\resizebox{0.9\columnwidth}{!}{\includegraphics{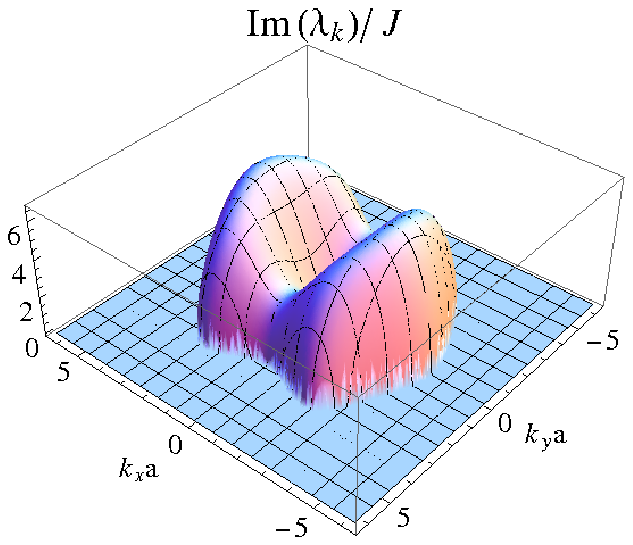}} 
\resizebox{0.9\columnwidth}{!}{\includegraphics{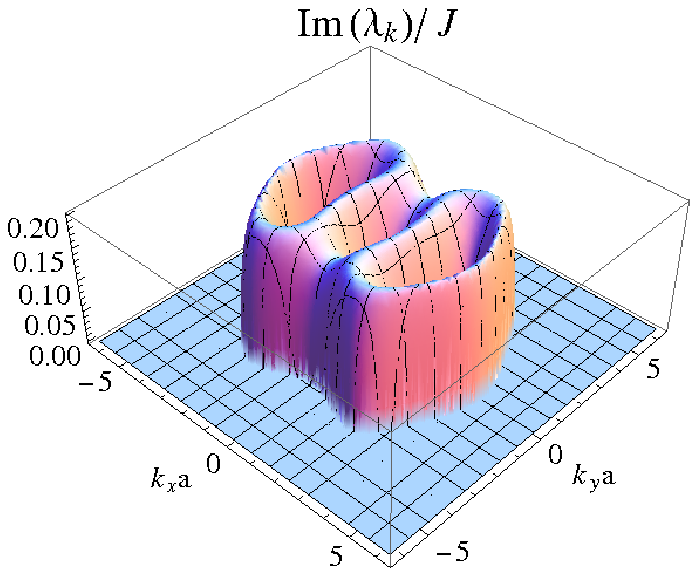}} 
\caption{(Color on line) Lyapunov exponent, i.e.  imaginary part of $\lambda_{\vect{k}}$, see Eq.~(\ref{lyapun}), in the units of bare tunneling amplitude $J$ as a function of  $k_x$ and $k_y$ for the ferromagnetic state and  for tunneling amplitudes in the regime of the S1$^\alpha$ phase. The third quasi-momentum coordinate $k_z$ corresponds to the maximum of ${\rm Im}(\lambda_{\vect{k}})$. Tunneling amplitudes are  $J_1=-0.2J$, $J_2= -0.1J$ and $J_3=0.2J$, while top (bottom) panel is related to $nU=50J$ ($nU=0.2J$). }
\label{bogol}
\end{center}
\end{figure}

A mean field analysis is a practical and convenient tool when describing weakly interacting cold bosonic gases in optical lattices. Stationary solutions of the Gross-Pitaevskii equation of a homogeneous system described by the Bose-Hubbard Hamiltonian are given in form of Bloch waves 
\be
\psi_0(\vect{r}_i)=\frac{1}{\sqrt{N_s}} \;e^{i\vect{k}_0 \cdot \vect{r}_i},
\label{blochw}
\ee 
where $\vect{k}_0$ is a quasi-momentum vector and $N_s$ stands for a number of lattice sites.
Even though we assume weak repulsive atomic interactions (i.e. $nU\ll J$ with mean number of $n$ atoms per lattice site) they may contribute to the dynamical instability of stationary states. For this reason, we perform stability analysis of mean-field solutions. 

The Gross-Pitaevskii equation corresponding to the Bose-Hubbard model (\ref{BHH}) can be obtained by switching from the operators to c-numbers, i.e. $\hat b_i\rightarrow b_i$, in the Heisenberg equation for $\hat b_i$ \cite{pethick}. Linearization of this equation leads to standard Bogoliubov-de~Gennes equations \cite{pethick}. In the present situation in order to find stationary Bogoliubov modes we employ the following ansatz
\be
\left[ \!
\begin{array}{c}
{u}_{\vect k}(\vect r_i) \\ v_{\vect k}(\vect r_i)
\end{array}
\! \right]=
\frac{e^{i\vect k\cdot \vect r_i}}{\sqrt{N_s}}\left[ \!
\begin{array}{c}
U_{\vect k}\;e^{i\vect k_0\cdot \vect r_i}\\ V_{\vect k}\;e^{-i\vect k_0\cdot \vect r_i}
\end{array}
\! \right].
\label{bdgsolut}
\ee
The resulting Bogoliubov-de~Gennes eigenvalue problem is represented by a block diagonal matrix in which diagonal elements are $2\times 2$ matrices. Diagonalization of the blocks gives the eigenvalues
\be\label{wwlasne}
\lambda_{\vect{k}}^{\pm}=
\frac{1}{2}\left( \Delta E_{\vect{k}} - \Delta \tilde E_{\vect{k}} \pm \delta \lambda_\vect{k} \right),
\ee
where
\bea
\delta\lambda_\vect{k} &=& \sqrt{(\Delta E_{\vect{k}} +\Delta \tilde E_{\vect{k}})(4nU+\Delta E_{\vect{k}} +\Delta \tilde E_{\vect{k}})}, \cr &&\cr
\Delta E_{\vect{k}} &=& E(\ka_{0}+{\vect{k}})-E(\ka_0), \cr &&\cr
\Delta \tilde E_{\vect{k}} &=& E(\ka_{0}-{\vect{k}})-E(\ka_0),
\eea
and $E(\vect{k})$ is the dispersion relation (\ref{disp}). When for a certain $\vect{k}$ vector, a quantity $(\delta \lambda_{\vect{k}})^2$ is negative, then the corresponding eigenvalues are complex. In such a case, a stationary state (\ref{blochw}) is not dynamically stable and tends to collapse into a dominant Bogoliubov mode, i.e. a mode related to an eigenvalue with the largest imaginary part.

The stability of excited states may be a relevant experimental problem, when one prepares the system in a ground state $\psi_0$ and then changes the tunneling amplitudes. If $\psi_0$ is not the ground state of the new Hamiltonian, it may collapse on time scale depending how strong the repulsive particle interactions are. 

Usually, it is the most convenient to prepare the triangular lattice system in the ferromagnetic state, i.e. in the Bloch wave (\ref{blochw}) with $\vect{k}_0=0$ \cite{struck2011}. Assume that after the preparation, experimentalists change abruptly the tunneling amplitudes switching to a regime of the spiral S1$^\alpha$ phase. If the initial state is the ferromagnetic one, the eigenvalue expression (\ref{wwlasne}) simplifies to $\lambda_\vect{k}^{\pm} = \pm\lambda_\vect{k}$ with
\be
\lambda_\vect{k}=\sqrt{\Delta E_{\vect{k}}\;(2nU+\Delta E_{\vect{k}})},
\label{lyapun}
\ee
where $\Delta E_{\vect{k}}=E(\vect{k})-E(0)$ is negative for some $\vect{k}$.
For $U=0$ the ferromagnetic state is still a stable stationary solution of the Gross-Pitaevskii equation. On the other hand, if $nU$ is larger than $\max(-\Delta E_{\vect{k}}/2)$, a small perturbation of the ferromagnetic wave-function will grow exponentially in time and occupations of the Bogoliubov modes with  $\vect{k}$ corresponding to the S1$^\alpha$ ground states increase the fastest. Figure~\ref{bogol} shows imaginary part of $\lambda_\vect{k}$ (the so-called Lyapunov exponent) where we can see that the maximal Lyapunov exponent is related to two quasi-momentum vectors. These vectors correspond to two degenerate ground states of the S1$^\alpha$ phase. We may expect that in an experiment one of the ground states of the S1$^\alpha$ phase will be predominantly populated --- which one depends on an initial perturbation and may be different in different experimental realizations. Signatures of such a spontaneous symmetry breaking have been observed experimentally in the 2D case \cite{struck2011}. 

In a very weak interaction regime, namely when $nU$ is smaller than $\max(-\Delta E_{\vect{k}}/2)$, a situation is different. Dominant modes can be found from the condition
\be
\label{warunek}
\Delta E_{\vect{k}}+nU=0.
\ee
An example of such a situation is presented in Fig.~\ref{bogol} where we can see a family of the most unstable modes. Interestingly the Bogoliubov modes with $\vect{k}$ corresponding to the S1$^\alpha$ ground states are stable. Thus, there is no decay of the ferromagnetic state towards the new ground states. 

The analyzed effects may by observed in experiments. For instance, we take realistic conditions  $nU=50J$ with $J=\hbar\times 6.5 $ Hz and calculate the stability of the ferromagnetic state in the region of the S1$^\alpha$ phase. For $J_1=-0.2J$, $J_2=-0.1J$ and $J_3=0.2J$,  the energy difference $\max (-\Delta E_\vect{k}/2) = 0.45J$  and as a result the ferromagnetic state should predominantly collapse towards one of the ground states on time scale of $1/\left(2|\lambda_\vect{k}|\right)\approx 11$ms. 
  On the other hand, if we consider smaller mean number of atoms per lattice site, e.g. $nU=0.2J$, we may expect the ferromagnetic state to decay predominately towards a multiple modes corresponding to the maximal Lyapunov exponent on time scales of $390$ms. 

%-------------------------------------------------------------------------------------------------------------------------------

\section{Complex tunneling amplitudes and time-reversal symmetry}
\label{timereversal}
%-------------------------------------------------------------------------------------------------------------------------------

%In this section we consider purely periodical motion of the lattice, neglecting a turn-on of the modulation, i.e. we abandon the Heaviside step function (\ref{veloc}). 

\begin{figure}
\begin{center}
\resizebox{0.8\columnwidth}{!}{\includegraphics{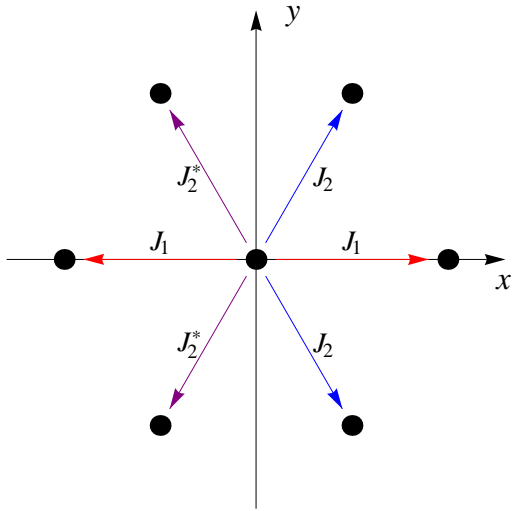}} 
\caption{(Color on line) Bravais lattice points (black circles) and amplitudes $J_i$ corresponding to tunneling from a lattice point to the nearest neighbors.}
\label{tunn}
\end{center}
\end{figure}

\begin{figure}
\begin{center}
\resizebox{0.8\columnwidth}{!}{\includegraphics{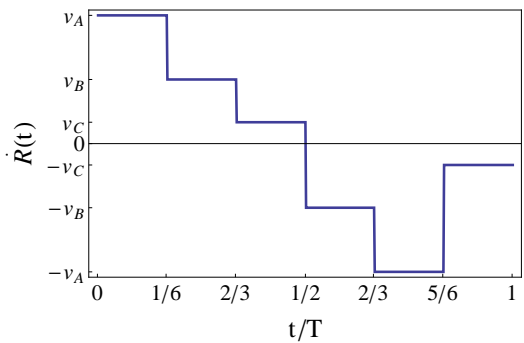}} 
%\resizebox{0.85\columnwidth}{!}{\includegraphics{F1D.eps}} 
\caption{An example of a velocity of a shaken 1D optical lattice with no time reversal symmetry that leads to an effective Bose-Hubbard model with real tunneling amplitudes, see Eq.~(\ref{exampl2}). }
\label{1d}
\end{center}
\end{figure}

In Sec.~\ref{secii} we have considered an optical lattice driven by a periodic external force and obtained an effective time-independent Bose-Hubbard Hamiltonian (\ref{BHH}) with real tunneling amplitudes (\ref{j1j2j3}). The original time-dependent Hamiltonian is invariant under a generalized time-reversal symmetry. In other words, there exists an anti-unitary operator $\hat{\mathcal{T}}$ which commutes with the Hamiltonian. In such a case, it can be easily shown how to construct a basis in which Hamiltonian is represented by a real symmetric matrix \cite{haake}. As a consequence,  a standard idea to obtain effective Bose-Hubbard models with complex tunneling amplitudes is to apply external driving that breaks that symmetry  \cite{TRSbreaking,struck2012,struck2012a}. 
However, we will show that an effective complex Bose-Hubbard model can also emerge even if a system is invariant under generalized time reversal transformation $\hat{\mathcal{T}}$. 

%{\blue We will also show that the absence of any generalized time-reversal invariance of a periodically driven lattice system is not a sufficient condition in order to obtain the effective Bose-Hubbard Hamiltonian with complex tunneling amplitudes. }

%We will also show an opposite example where an effective Bose-Hubbard Hamiltonian is represented by a real matrix even though {\blue a system does not possess any generalized time-reversal symmetry.}

For the sake of simplicity, let us consider a 2D bosonic gas in a triangular optical lattice with primitive vectors $\vect{a}_1$ and $\vect{a}_2$, see Eq.~(\ref{a1a2a3}), subjected to a periodic driving that results in a motion of the lattice on a periodic trajectory with a velocity
\be
\label{velocity2D}
\begin{split}
\vect{\dot{R}}(t)&= \frac{F_x}{m\omega}\vect{e}_x \cos (\omega t) -\frac{F_y}{m\omega}\vect{e}_y \sin(\omega t)+ \\
 &-\frac{\tilde{F}_y}{2 m\omega}\vect{e}_y \sin(2\omega t).
\end{split}
\ee
The Hamiltonian (\ref{bosehubbard}), where $\vect{F}(t)=-m\vect{\ddot{R}}(t)$, does not change under the generalized time-reversal transformation   $\hat{\mathcal{T}} =\hat P_x\hat T$, where $\hat T$ is the time-reversal operation and $\hat P_x$ stands for $x\rightarrow -x$. \cite{haake}. The Wannier basis vectors are not invariant under the $\hat P_x\hat T$ transformation and consequently the Floquet Hamiltonian is not necessarily represented by a real symmetric matrix in that basis. Thus, we may expect that the effective Bose-Hubbard Hamiltonian (\ref{BHH}), which is a single block of the entire Floquet Hamiltonian matrix \cite{eckardt2005,TRSbreaking},  may possess complex tunneling amplitudes. Indeed, the effective tunneling amplitude
\bea
J_1&=&\frac{J}{T}\int_0^T dt\; \exp\left(\frac{i\,m}{\hbar}\vect{a}_1\cdot \dot{\vect{R}}(t)\right)\cr && \cr
&=&J\;{\cal J}_0\left(\frac{F_xa}{\hbar \omega}\right), 
\eea
is real but 
\bea
J_2&=&\frac{J}{T}\int_0^T dt\; \exp\left(\frac{i\,m}{\hbar}\vect{a}_2\cdot \dot{\vect{R}}(t)\right)\cr && \cr 
&=&J \sum_{n=-\infty}^{+\infty} \;{\cal J}_{2n}\left(\frac{K_1}{\hbar \omega}\right){\cal J}_{n}\left(\frac{K_2}{2\hbar \omega}\right) e^{-i2n \zeta},
\label{j1j2}
\eea
is complex valued if $\tilde F_y\ne0$ and $\zeta=F_x/ \left( \sqrt{3} F_y \right)$ is not a multiple of $\pi/2$. In Eq.~(\ref{j1j2}) $K_1 = a\sqrt{F_x^2+3F_y^2}/2$ and $K_2= a \sqrt{3}  \tilde{F}_y/2 $. Amplitudes related to tunneling in different directions are indicated in Fig.~\ref{tunn}. This figure shows also that the generalized time-reversal symmetry is preserved by the effective Hamiltonian, i.e. complex conjugation combined with reflection $x\rightarrow -x$ does not change the Hamiltonian.

Now let us consider a problem if the absence of any generalized time-reversal invariance of a periodically driven lattice system is a sufficient condition in order to obtain the effective Bose-Hubbard Hamiltonian with complex tunneling amplitudes? We will see that it is not, i.e. we are able to construct an example where any generalized time-reversal  symmetry is broken but tunneling amplitudes, in an effective Bose-Hubbard Hamiltonian, are real valued. Consider a simple 1D model, driven with a periodic piecewise constant velocity 
\be
\dot{R}(t)=\left\{
\begin{array}{rc}
v_A ,& 0<t < T_1\cr
v_B ,&  T_1<t <2T_1\cr
v_C ,&   \cr
-v_B ,& \ldots \cr
-v_A ,&  \cr
-v_C ,&  5T_1<t <6T_1\cr
\end{array}
%\mod T,
\right.,
\label{exampl2}
\ee
where $T_1=T/6$ and 
%$|v_A|\ne|v_B|\ne|v_C|$ and $|v_A|\ne|v_C|$,
$v_A>v_B>v_C>0$,
 see Fig.~\ref{1d}. It is easy to check that the resulting tunneling amplitude is real valued,
\bea
J_1&=&\frac{J}{T}\int_0^T dt\; \exp\left(\frac{im}{\hbar}\dot{R}(t)\right) 
\cr &=& 
\frac{J}{3}\left[ \cos \left(\frac{mv_A}{\hbar}\right)+\cos \left(\frac{mv_B}{\hbar}\right)+\cos \left(\frac{mv_C}{\hbar}\right)\right].
\cr &&
\eea 
Effective Bose-Hubbard Hamiltonian is a single block of the entire Floquet Hamiltonian of a shaken lattice system \cite{eckardt2005,TRSbreaking}. We see that such a block can be real even if the entire matrix is expected to be complex Hermitian. 

In Ref.~\cite{struck2012} two conditions have been  considered in order to achieve an effective complex Bose-Hubbard model. That is, a tunneling amplitude between two lattice sites $\vect{r}_i$ and $\vect{r}_j$ can be complex, if the projection of an inertial force $f_{ij}(t)= (\vect{r}_i-\vect{r}_j)\cdot\vect{F}(t)$  breaks: (a) reflection symmetry for suitable time $\tau$, i.e. $f_{ij}(t-\tau)=f_{ij}(-t-\tau)$ and (b) shift anti-symmetry, i.e. $f_{ij}(t-\pi/\omega)=-f_{ij}(t)$.  The model presented in Fig.~\ref{1d} consitutes an example where the conditions are broken but the tunneling amplitudes remain real valued. 

 %-------------------------------------------------------------------------------------------------------------------------------
 %-------------------------------------------------------------------------------------------------------------------------------
\section{Conclusions}
\label{concl}
%-------------------------------------------------------------------------------------------------------------------------------
In this paper we propose a complete realization of a quantum simulator of the classical XY model in ultra-cold atoms in a three-dimensional triangular optical lattice and present its predictions. In order to manipulate independently values of three different couplings we consider a motion of the lattice constrained to an ellipse. 

Out of simulations we obtain a wide variety of spin configurations, some of which being generalizations of the two-dimensional triangular lattice model, and a full phase diagram. In particular we observe the emergence of a new phase, enabling a second order phase transition between ferromagnetic and rhomboidal phases, and a discontinuous phase transition between two types of spiral phases. This discontinuous phase transition is a consequence of $J_3$ interaction, absent in the two-dimensional model. 

Optical lattice experiments start usually with ferromagnetic state and then parameters of the lattice are slowly or suddenly changed in order to switch to a different phase. We have applied Bogoliubov approach which allows us to analyze stability of stationary mean-field solutions under a change of system parameters. For sufficiently strong particle interactions the initial ferromagnetic state collapses towards a ground state --- for typical experimental parameters it takes place during a few millisecond. On the other hand if the interactions are too weak the initial state loses its stability but it does not evolve towards a system ground state.

In the present paper we mostly concentrate on a Bose-Hubbard model with real tunneling amplitudes. However, we consider also a problem of the realization of complex tunneling amplitudes by means of periodic shaking of an optical lattice. It is known that a time periodic perturbation that the breaks time-reversal symmetry leads to a Floquet Hamiltonian that is represented by a complex Hermitian matrix and the effective Bose-Hubbard model can be expected to possess complex tunneling amplitudes. However, we show that it is also possible to realize such amplitudes with a perturbation that is invariant under a time-reversal transformation.

%-------------------------------------------------------------------------------------------------------------------------------
\section*{Acknowledgments}

Support of Polish National Science Center via project DEC-2012/04/A/ST2/00088 is acknowledged.
%-------------------------------------------------------------------------------------------------------------------------------

%%%%%%%%%%%%%%%%%%%%%%%%%%%%%%%%%%%%%%%%%%%%%%%%%%%%%%%%%%%%%%%%%%%%%%%%%%%

%%%%%%%%%%%%%%%%%%%%%%%%%%%%%%%%%%%%%%%%%%%%%%%%%%%%%%%%%%%%%%%%%%%%%%%%%%%%

\end{document}